\documentclass{ws-ijmpa}

\usepackage[super]{cite}
\usepackage{graphicx}
\usepackage{amssymb}
\usepackage{amsmath}
\usepackage[svgnames]{xcolor}
\usepackage{mathtools,slashed}
\usepackage[retainorgcmds]{IEEEtrantools}
\usepackage{physics}
\usepackage{epstopdf}
\usepackage[utf8]{inputenc}
\usepackage{url}
\usepackage{subfig}
\usepackage[colorlinks,citecolor=DarkGreen,linkcolor=DarkRed,urlcolor=DarkBlue]{hyperref}
\usepackage{orcidlink}
\usepackage{hyperref}

\begin{document}

\markboth{A. Ayala et al.}{Searching for signatures of criticality in heavy-ion collisions}

%
\catchline{}{}{}{}{}
%

\title{Searching for signatures of criticality in heavy-ion collisions using two-pion correlation functions
}

\author{\mbox{Alejandro Ayala~\orcidlink{0000-0003-3929-9209}~$^{1,2,3}$}, \mbox{Santiago Bernal-Langarica~\orcidlink{0000-0003-0826-9349} \footnote{email: \href{mailto:santiago.bernal@correo.nucleares.unam.mx}{santiago.bernal@correo.nucleares.unam.mx} (corresponding author)}~$^{1}$}, \mbox{Máté Csanád\,\orcidlink{0000-0002-3154-6925}\,$^{4}$}, \mbox{Isabel Domínguez~\orcidlink{0000-0001-6831-3159}${}^{5}$}, \mbox{Irving I. Gaspar~${}^{6}$}, \mbox{Luis A. Hernández~\orcidlink{0000-0001-8764-835X}~${}^{6}$}, \mbox{Johannes Jahan~\orcidlink{0000-0002-4557-4652}~${}^{7}$}, \mbox{Micheal Kahangirwe~\orcidlink{0000-0001-9144-6240}~${}^{8}$}, \mbox{Jamie M. Karthein~\orcidlink{0000-0003-2041-5206}~${}^{9}$}, \mbox{Dániel Kincses~\orcidlink{0000-0002-8880-6732}~${}^{4}$}, \mbox{Grégoire Pihan~\orcidlink{0000-0002-7072-4480}~${}^{7}$}, \mbox{Tomáš Poledníček~\orcidlink{0009-0009-8471-3051}~${}^{10}$}, \mbox{María Elena Tejeda-Yeomans~\orcidlink{0000-0002-4938-7563}~$^{11}$} and \mbox{Boris Tomášik~\orcidlink{0000-0003-0478-0609}~${}^{10,12}$}}
\address{${}^{1}$ Instituto de Ciencias Nucleares, Universidad Nacional Aut\'onoma de M\'exico, Apartado Postal 70-543, Ciudad de México 04510,
  Mexico.}
\address{${}^{2}$ Instituto de Física, Universidade de São Paulo, Rua do Matão, 1371, CEP 05508-090, São Paulo, SP, Brazil}
\address{${}^{3}$ Instituto de Física Teórica, Universidade Estadual Paulista, Rua Dr. Bento Teobaldo Ferraz, 271 - Bloco II, 01140-070 São Paulo, SP, Brazil}
\address{${}^{4}$ ELTE Eötvös Loránd University, Institute of Physics and Astronomy, Pázmány Péter sétány 1/a, Budapest, H-1117, Hungary}
\address{${}^5$ Facultad de Ciencias Físico-Matemáticas, Universidad Autónoma de Sinaloa,
Avenida de las Américas y Boulevard Universitarios, Ciudad Universitaria, Culiacán, 80000, Mexico}
\address{${}^{6}$ Departamento de Física, Universidad Autónoma Metropolitana-Iztapalapa,
Avenida San Rafael Atlixco 186, Ciudad de México 09340, Mexico}
\address{${}^{7}$ Department of Physics, University of Houston, Houston, TX 77204, USA}
\address{${}^{8}$ Center for Nuclear Research, Department of Physics, Kent State University, Kent, 44242, OH, USA}
\address{${}^{9}$ Cyclotron Institute and Department of Physics and Astronomy, Texas A\&M University, College Station, TX 77845, USA}
\address{${}^{10}$ Fakulta jaderná a fyzikáln\v e in\v zen\'yrská, \v Ceské vysoké u\v cení technické v Praze, B\v rehová 7, 11519 Praha 1, Czech Republic}
\address{${}^{11}$ Facultad de Ciencias-CUICBAS, Universidad de Colima,
Bernal Díaz del Castillo No. 340, Colonia Villas San Sebastián, Colima, 28045, Mexico}
\address{${}^{12}$ Univerzita Mateja Bela, Tajovského 40, 974 01 Banská Bystrica, Slovakia}

\maketitle

\begin{history}
\received{Day Month Year}
\revised{Day Month Year}
\accepted{Day Month Year}
\published{Day Month Year}
\end{history}

\begin{abstract}

We study possible signatures of the existence of a Quantum Chromodynamics (QCD) critical end-point in low-energy relativistic heavy-ion collisions using two-pion femtoscopy. We simulate central $^{209}\mathrm{Bi}+^{209}\mathrm{Bi}$ collisions at center-of-mass energies ranging from $\sqrt{s_{NN}}=5.8$ to 11 GeV using the EPOS 4.0.3 framework, incorporating an equation of state that explicitly includes a critical end-point and is compatible with lattice QCD results at small baryon chemical potentials. We analyze the pion pair-distance distribution and the corresponding two-pion correlation function by fitting them with Lévy distributions in configuration and momentum spaces, respectively. We find a weak dependence of the source parameters on the collision energy below $\sqrt{s_{NN}}=9.2$ GeV, followed by a change in trend at higher energies, particularly in the intercept parameter $\lambda$. This behavior may be associated with a decreasing fraction of pions coming from the core and an increasing contribution from long-lived resonance decays. Within the present analysis, we do not identify an unambiguous signal attributable to the critical end-point. A comparison with simulations employing different equation of state parameters to vary the strength of the critical point, as well as with simulations without a critical end-point, is necessary to establish the sensitivity of these observables to critical phenomena.

\end{abstract}

\keywords{Heavy-ion collisions; femtoscopy; critical phenomena.}

\section{Introduction}	

Low-energy relativistic heavy-ion collisions are a prime tool to study the properties of strongly interacting matter in the regime of high baryonic densities and low temperatures. In particular, these systems can be used to probe the regions of the phase diagram of Quantum Chromodynamics (QCD) in which neither Lattice QCD (LQCD) nor perturbative QCD are valid. Also, these systems can be used to shed light on the existence and location of a critical end-point (CEP) in the phase diagram, as well as on the properties of the QCD equation of state (EoS). 

One of the techniques that has been developed in the past decades to study heavy-ion collisions is femtoscopy, since it can be used to determine the size, lifetime, and structure of the system created in such collisions~\cite{Boal:1990yh, Weiner:1999th, Lednicky:2005af, Lisa:2005dd, Csorgo:2005gd, Lisa:2008gf, Kisiel:2011jt, Lokos:2022exf, Janik:2018ghw, Ayala:2021zst, Ayala:2023sbb, Ayala:2001pf, Kincses:2025iaf}. Due to the fact that charged pions are the most abundant particle species produced in heavy-ion collisions, most of the femtoscopy studies have focused on the two-pion correlation function, which is described in terms of a set of parameters that gives information about the space-time structure of the pion-emitting source.

In particular, it has been proposed that the so-called Lévy index of stability, which is an accessible quantity in femtoscopic analysis, can be related to the critical exponent $\eta$ of QCD~\cite{Csorgo:2008ayr, Csanad:2020xbf, Csorgo:2005it}. In the case of the 3D Ising model universality class, the critical exponent $\eta$ is expected to be approximately $\eta \approx 0.04$, whereas for the Random-Field Ising model, it is approximately $\eta \approx 0.5$~\cite{Rieger:1995abc, Fytas:2013mrim}. Thus, the Lévy index of stability is expected to reach a minimum at the CEP and so it could serve as a possible signal of criticality. Previous studies~\cite{Ayala:2023sbb, Huang:2025edi, Porfy:2025sev} have explored this idea using Monte Carlo simulations but without using an EoS that includes a CEP. In this work we address this gap by considering an EoS that extends to finite baryon density, contains a CEP, and is compatible with LQCD results at small baryonic chemical potential. This work is organized as follows: in Sec.~\ref{sec:EoS} we present the equation of state at finite baryonic density and compatible with the LQCD results at small baryonic chemical potential, that will be used throughout this work. In Sec.~\ref{sec:dist} we present the analysis method of the two-pion correlation function and the pair-distance distribution in configuration space and in momentum space. In Sec.~\ref{sec:results} we show our results and compare the employed methods with each other. We finally summarize and conclude in Sec.~\ref{sec:conc}.

\section{Finite density QCD equation of state}\label{sec:EoS}

One family of QCD equations of state, defined in the $(T,\mu_B)$-plane, that is consistent with the LQCD results for small baryonic chemical potential is the one developed in Ref.~\refcite{Kahangirwe:2024cny}. This family of EoS is based on the assumption that the QCD critical point belongs to the 3D Ising model universality class~\cite{Csorgo:2008ayr, Stephanov:1998dy}. Here, we present a short review on the construction of this EoS, that is then used throughout this work and is denoted as the 2D-TExS EoS (two-dimensional $T'$-Expansion Scheme). In the crossover region of the QCD phase diagram, it was observed by the Wuppertal-Budapest lattice QCD Collaboration~\cite{Borsanyi:2021sxv} that the scaled baryonic density, as a function of the temperature, does not significantly change as a function of the imaginary scaled baryonic chemical potential. This defines a ``rescaled temperature'', $T'(T,\mu_B)$ such that
\begin{equation}
\label{eq:Tprime}
    T \frac{\chi_1 ^B (T, \mu_B)}{\mu_B} = \chi_2 ^B (T',0),
\end{equation}
where the reduced baryon number density, $\chi_1 ^B = \flatfrac{n_B}{T^3}$, at finite $\mu_B$ is obtained from the second-order susceptibility, $\chi_2 ^B$, calculated by LQCD at $\mu_B = 0$, defined like all other even-order baryon susceptibilities as
\begin{equation}
    \chi_n ^B (T) = \left.\pdv[n]{\left(\flatfrac{\mu_B}{T}\right)} \frac{P(T)}{T^4}\right|_{\mu_B = 0}\,.
\end{equation}
At finite baryonic chemical potential, the rescaled temperature $T'$ function can be expanded in powers of $\flatfrac{\mu_B}{T}$ as
\begin{equation}
\label{eq:Texp}
    T'(T,\mu_B) = T\left[ 1 + \kappa_2 ^{BB} (T) \left(\frac{\mu_B}{T}\right)^2 + \kappa_4 ^{BB} (T) \left(\frac{\mu_B}{T}\right)^4 + \ldots\right],
\end{equation}
where the coefficients $\kappa_{2n} ^{BB}$ are obtained from the LQCD susceptibilities. To include the critical behavior of the universality class of the 3D-Ising model, the reduced temperature, $r$, and the external magnetic field, $h$, of the Ising model are mapped into the QCD phase diagram variables under the non-universal transformation
\begin{IEEEeqnarray}{rCl}
    \frac{T' - T_0}{T_C T'_{,T}} & = & - w' h \sin\alpha'_{12}, \nonumber \\
    \frac{\mu_B ^2 - \mu_{BC}^2}{2\mu_{BC}T_C} & = & w' \left( -r\rho' -h \cos\alpha'_{12}\right),
\end{IEEEeqnarray}
where $T'_{,T} \equiv \left(\flatfrac{\partial T'}{\partial T}\right)_{\mu_B}$, $T_0$ is the transition temperature at $\mu_B = 0$, $T_C$ and $\mu_{BC}$ are the temperature and baryonic chemical potential of the CEP, and the free parameters $w'$, $\rho'$, and $\alpha'_{12}$ act as scaling factors for $r$ and $h$. These free parameters can be rewritten in terms of the free parameters of Ref.~\refcite{Parotto:2018pwx} as
\begin{IEEEeqnarray}{rCl}
    \tan\alpha_2 & = & \tan\alpha_1 - \tan\alpha'_{12}, \\
    w & = & w'\cos\alpha'_{12} \sqrt{1 + \left(\tan\alpha_1 - \tan\alpha'_{12}\right)^2}, \\
    \rho & = & \frac{\rho'}{\cos\alpha_1 \cos\alpha'_{12}\sqrt{1+ \left(\tan\alpha_1 - \tan\alpha'_{12}\right)^2}} .
\end{IEEEeqnarray}

Since $\chi_2 ^B (T',0)$ is an analytic function, the singularity associated with the critical point should be in the rescaled temperature $T'(T, \mu_B)$. Consequently, both $T'$ and the baryonic density contain a regular and a critical component, which are related to each other. In order to match the data from LQCD with the critical contribution, the $T'$ can be constructed as
\begin{equation}
    T'(T,\mu_B) = T'_\text{LQCD}(T,\mu_B) + T'_\text{crit}(T,\mu_B) - \text{Taylor}_{n\leq 2} \left[T'_\text{crit} (T, \mu_B)\right],
\end{equation}
where $T'_\text{LQCD}$ is given by Eq.~\eqref{eq:Texp}, truncated at second order in $\flatfrac{\mu_B}{T}$, and the last term in the previous equation represents the Taylor expansion of the critical contribution of $T'_\text{crit}(T,\mu_B)$, up to second order in $\flatfrac{\mu_B}{T}$. Then, by substituting the $T'$ given by the previous equation into Eq.~\eqref{eq:Tprime}, it is possible to compute the baryonic density for a given set of the free parameters. From that point, all relevant thermodynamic quantities can be obtained within the range \mbox{$0\leq \mu_B \leq 700$ MeV} and \mbox{$25 \leq T \leq 800$ MeV}.

In this work, we constructed such an EoS by using the following parameters: $\mu_{BC} = 400$ MeV, $T_C = 133.36$ MeV, $w = 4.25$, $\rho = 1$, and $\alpha_{12} \equiv \alpha_2 - \alpha_1 = \flatfrac{\pi}{2}$. We use the parametrization of Cleymans \textit{et al.}~\cite{Cleymans:2005xv} to estimate the collision energy corresponding to freeze-out parameters closest to our selected CEP position. This corresponds to a collision energy of $\sqrt{s_{NN}} = 8.3$ GeV.

\section{Two-pion correlation function and pair-distance distribution}\label{sec:dist}

When working with Monte Carlo simulations, such as those obtained from EPOS~\cite{Werner:2023fne, Werner:2023jps, Werner:2023mod, Werner:2023zvo, Werner:2024fwk, Werner:2024ntd, Jahan:2024ycu}, the charged pion pair-distance distribution can be obtained directly~\cite{Huang:2025edi, Molnar:2026vgi, Kincses:2025izu, Kincses:2026tfa, Porfy:2025sev}, in contrast to an experimental case, where the freeze-out coordinates of the particles are unknown. This charged pion pair-distance distribution, $D$, is related to the two-pion correlation function, $C_2$ through
\begin{equation}
    C_2(q, K) = \int \dd[4]{r} D(r,K) \left|\Psi_q ^{(2)} (r) \right|^2,
\end{equation}
where $q = p_1 - p_2$ is the pair relative momentum, $K$ is the pair average momentum, $r$ the pair separation and $\Psi_q ^{(2)}$ is the pair wavefunction. $\left|\Psi_q ^{(2)} (r)\right|^2 = 1 + \cos(q\cdot r)$ contains the information about the Bose-Einstein correlations of the pair and neglects final-state interactions~\cite{Kincses:2019rug, Nagy:2023zbg}; however, throughout this work, we will ignore the latter. Both the two-pion correlation function and the pair-distance distribution contain the space-time information that characterizes the particle emitting source. Furthermore, after imposing the on-shell condition for the pair, $C_2$ depends only on the three spatial components of $\vec{q}$ and the three spatial components of $\vec{K}$.

It has been argued that the two-pion correlation function, and therefore the pair-distance distribution, is better described by a Lévy distribution, rather than a Gaussian~\cite{Kincses:2024lnv, Csorgo:2003uv, Cimerman:2019hva, Ayala:2023sbb, Arpasi:2024ohw, NA61SHINE:2023qzr, PHENIX:2017ino, PHENIX:2024vjp}. By fitting the data obtained from the Monte Carlo simulations to a Lévy distribution it is possible to obtain the parameters that characterize the particle emitting source in terms of the source radii, the intercept parameter, and the Lévy index of stability. In the following, we will describe how this process is done in both configuration space and momentum space.

In order to perform the analysis of the charged pion pair-distance distribution, it is necessary to choose a reference frame. A common choice is the longitudinally comoving system (LCMS) in the Bertsch-Pratt coordinates~\cite{Bertsch:1988db, Pratt:1986cc}, which has the advantage that, for mid-rapidity, the source is approximately spherical for RHIC energies~\cite{PHENIX:2017ino}. The LCMS is the frame, boosted along the beam axis, in which the longitudinal component of the pair average momentum vanishes ($K_z = 0$); the longitudinal direction is along the beam, the outward direction along the pair transverse momentum, $K_T$, and the sideward direction perpendicular to both.
For a vector, $\vec{\rho}$, in that coordinate system its components can be computed from the laboratory coordinates as~\cite{Kincses:2024lnv}
\begin{IEEEeqnarray}{rCl}
\rho_\text{out} ^\text{LCMS} & = & r_x \cos\phi + r_y \sin\phi - \frac{K_T}{K_0 ^2 - K_z ^2} \left(K_0 t - K_z r_z\right), \nonumber\\
\rho_\text{side} ^\text{LCMS} & = & -r_x\sin\phi + r_y \cos\phi, \nonumber\\
\rho_\text{long} ^\text{LCMS} & = & \frac{K_0 r_z - K_z t}{\sqrt{K_0 ^2 - K_z ^2}},
\end{IEEEeqnarray}
where $\cos\phi = \flatfrac{K_x}{K_T}$, $\sin\phi = \flatfrac{K_y}{K_T}$, $K_T = \sqrt{K_x ^2 + K_y ^2}$ is the pair transverse momentum and ($r_x, r_y, r_z, t$) are the components of the pair separation in the laboratory frame.

In this work, we studied ${}^{209}\mathrm{Bi}+{}^{209}\mathrm{Bi}$ collisions within the EPOS 4.0.3 framework in the collision energy regime around the CEP described in Sec.~\ref{sec:EoS}, that is, \mbox{$\sqrt{s_{NN}} = 5.8 - 11$ GeV}. We study the same collision energy interval as in Ref.~\refcite{Ayala:2023sbb}, considering the 0--5\% centrality class. We chose 5000 full events, corresponding to five freeze-out events per hydrodynamical event, as the default value for the analysis and explored the dependence on the number of events, as well as the dependence on the pair transverse mass by dividing it into ten transverse mass intervals, $m_T = \sqrt{m^2 + K_T^2}$, with $m$ the pion mass and $m_T \in [0.2, 0.7]$ GeV, since in that interval, the systematic uncertainties do not depend on $m_T$~\cite{Huang:2025edi}. We focused on charged pions coming from the pseudo-rapidity region of $|\eta| \leq 1$, pions whose transverse momentum $p_T \in [0.15, 1]$ GeV, and pairs whose relative momentum in the LCMS is bounded by $q_\text{LCMS} ^\text{max} < c\cdot \sqrt{m_T \cdot 0.001 \text{ GeV}}$, with $c$ a dimensionless quantity. This simulates the typical kinematic cuts of the STAR detector~\cite{Korodi:2022ohn}. In this work, the statistical uncertainties are, in general, negligible. However, there are different important sources of systematic uncertainties. The different sources of systematic uncertainty are discussed in Secs.~\ref{sec:drho} and~\ref{sec:cq}. Here, we focus on the number of events ($N_\text{events}$), $q_\text{LCMS}^\text{max}$, and the fit range. To compute the total effect of the systematic uncertainties, we follow the analysis of Ref.~\refcite{Porfy:2025sev},
\begin{IEEEeqnarray}{rCl}
    \sigma_{i,j} ^2 [P] & = & \left(P^\text{def} - P_i ^\text{var}\right)^2 , \\
    \sigma_j ^\text{tot} [P] & = & \sqrt{\sum_{i = 1} ^n \sigma_{i,j} ^2 [P]}
\end{IEEEeqnarray}
where $P^\text{def}$ is the default value of a parameter, $P_i ^\text{var}$ its varied value for the $i$-th source of uncertainty, $n$ is the number of uncertainty sources ($n = 3$ in this case) and $j = +,-$ refers to the (possibly) asymmetric positive or negative uncertainty.

\subsection{Configuration space method}\label{sec:drho}

In configuration space, a Lévy distribution can be written as~\cite{Csorgo:2003uv}
\begin{equation}
\label{eq:Drho-levy}
    \mathcal{L} (\alpha, R^2, \vec{r}) = \frac{1}{(2\pi)^3} \int \dd[3]{\omega} e^{i\vec{r}\cdot\vec{\omega}} e^{-\frac{1}{2}\left|\vec{\omega}^T R^2 \vec{\omega}\right|^{\flatfrac{\alpha}{2}}},
\end{equation}
where $\alpha$ is the Lévy stability index, $R^2 = \text{diag}(R_\text{out} ^2, R_\text{side} ^2, R_\text{long} ^2)$ is the matrix of radii, $\vec{r}$ are the spatial coordinates and $\vec{\omega}$ is an integration variable. We then fit the pair-distance distribution using $\lambda \cdot \mathcal{L}(\alpha, R^2)$, where $\lambda$ is the intercept parameter. Hence, there are five different parameters to be fitted: $\lambda$, $\alpha$, $R_\text{out}$, $R_\text{side}$ and $R_\text{long}$. Since the Lévy distribution is non-analytic, its  implementation in numerical data fitting introduces significant errors associated with the numerical integration. To solve this issue, we take one-dimensional projections onto the longitudinal, outward and sideward directions and fit them simultaneously to one-dimensional projections of Eq.~\eqref{eq:Drho-levy}~\cite{Kincses:2024lnv, Csanad:LevyFit}.

As discussed previously, we consider as sources of systematic error the number of analyzed events, $N_\text{events}$, $q_\text{LCMS} ^\text{max}$ and the fitting interval. Here we consider $N_\text{events} ^\text{def} = $ 5000 events as the default value for the number of analyzed events and explored the effect for $N_\text{events} =$ 2000, 4000, 5000, 7500 and 10000 as the different settings. For $q_\text{LCMS} ^\text{max} = c\cdot \sqrt{m_T \cdot 0.001 \: \text{GeV}}$ we take $c^\text{def} = 12$ as the default value and explored the effect for $c = 8$, 10, 12, 15 and 18. The fit is made for $\rho \in [1\text{ fm}, \rho_\text{max} ^\text{fit}]$, with $\rho_\text{max} ^\text{fit} = 15$, 18, 20, 25 and 30 fm, with $\rho_\text{max} ^\text{fit, def} = 20$ fm, for all three directions.

As a function of $N_\text{events}$, $D(\rho)$ becomes almost flat for the considered $N_\text{events}$ values. As a function of $c$, $D(\rho)$ changes less than 4\%, while as a function of $\rho_\text{max}^\text{fit}$, $D(\rho)$ is also flat for $\rho_\text{max} ^\text{fit} \gtrsim 18$ fm.

\begin{figure}
    \centering
    \subfloat[Out]{\includegraphics[width=0.333\linewidth]{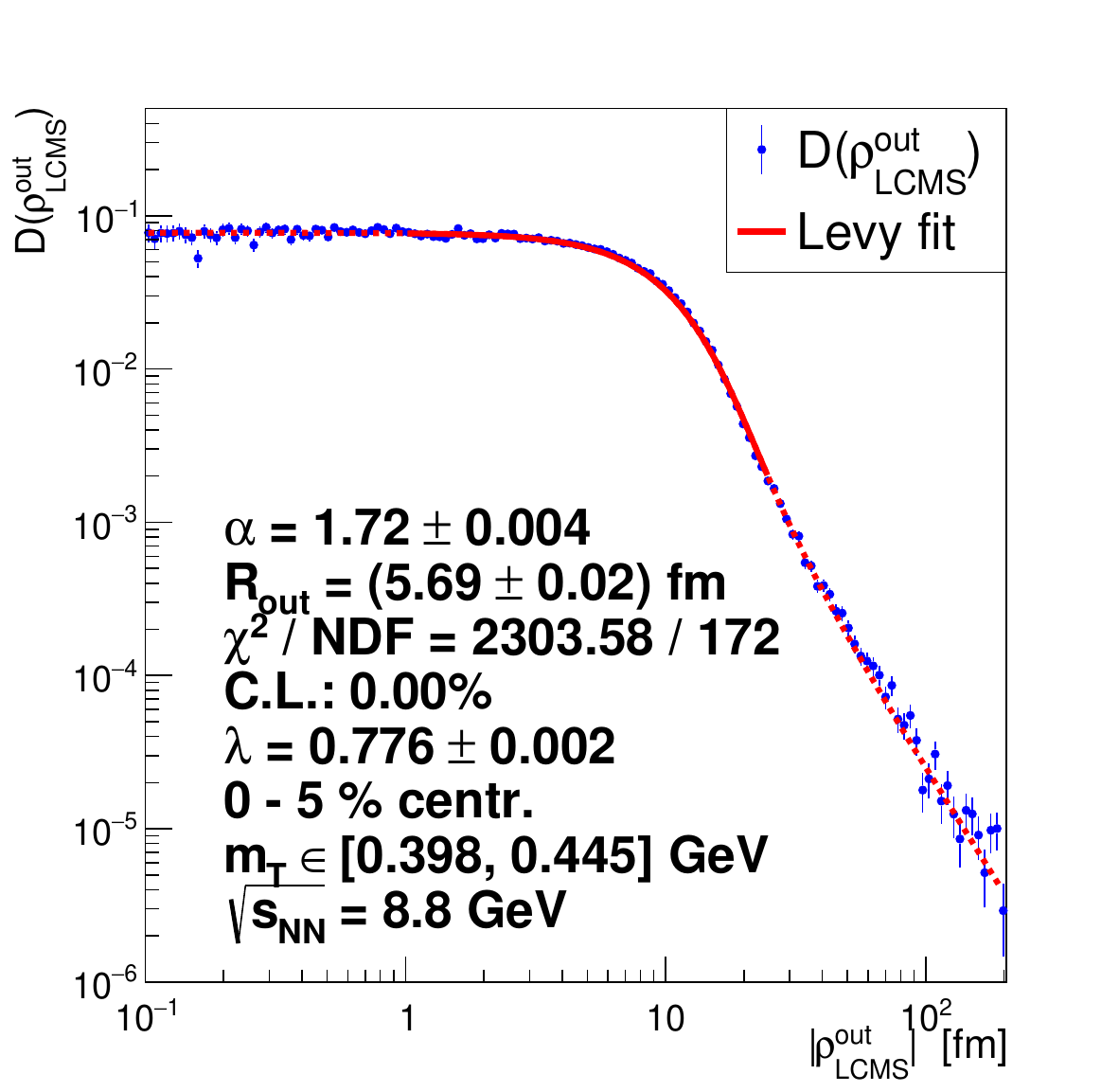}}
    \subfloat[Side]{\includegraphics[width=0.333\linewidth]{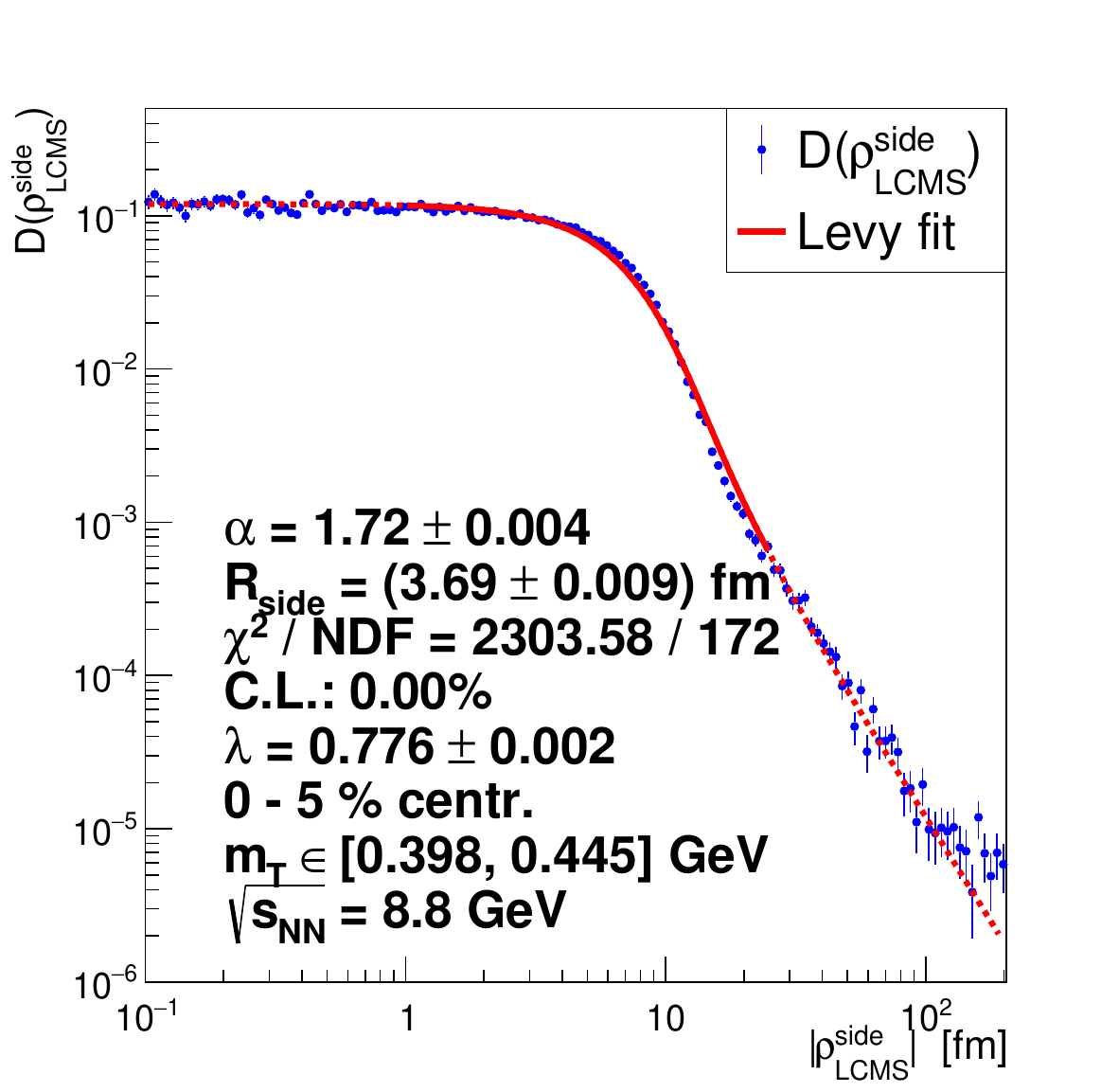}}
    \subfloat[Long]{\includegraphics[width=0.333\linewidth]{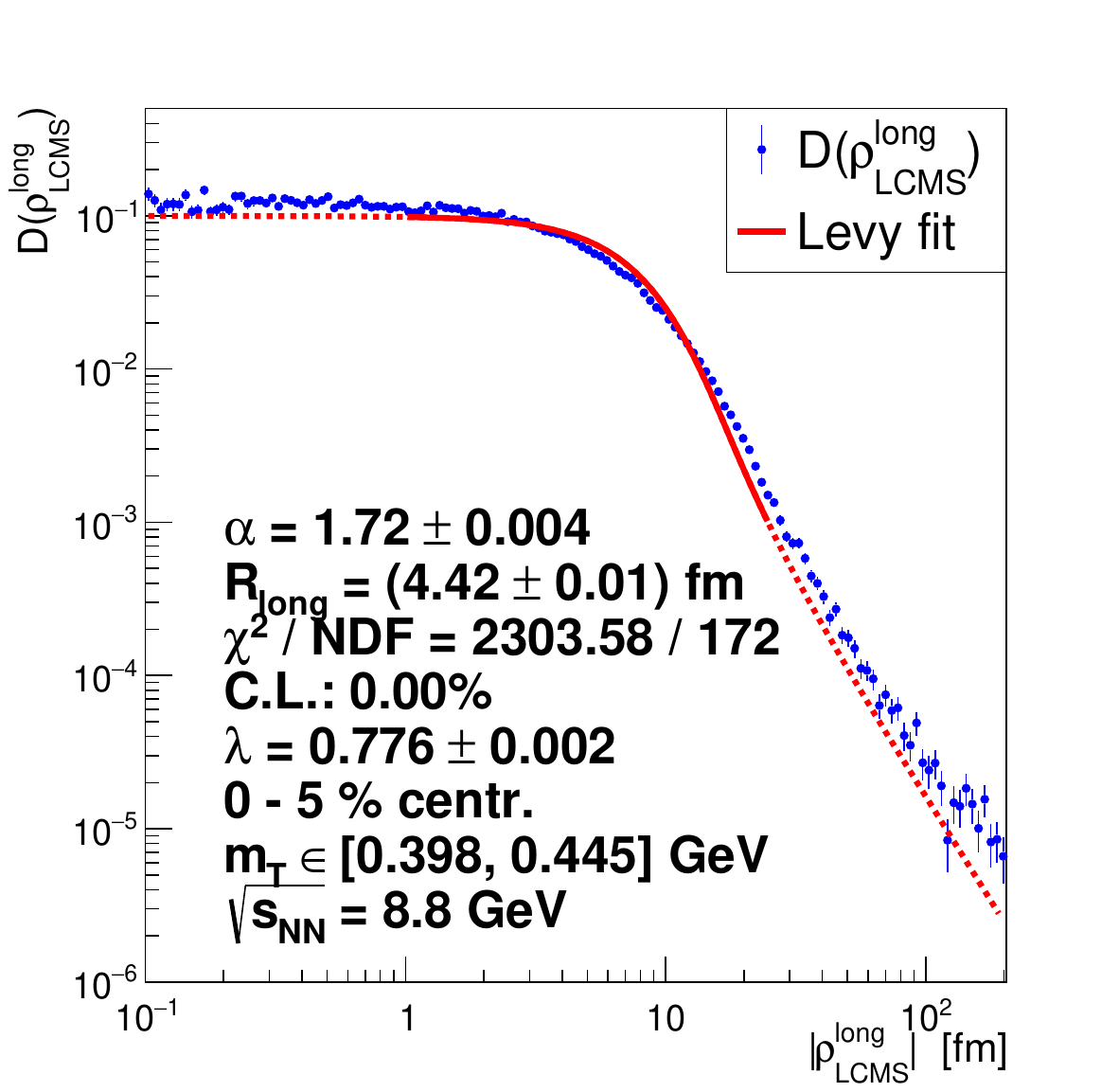}}
    \caption{Examples of the simultaneous fits to the pair-distance distribution along the (a) outward, (b) sideward and (c) longitudinal directions, obtained from ${}^{209}\mathrm{Bi}+{}^{209}\mathrm{Bi}$ collisions within the EPOS 4.0.3 framework with the 2D-TExS EoS. The blue dots represent the pair-distance distribution, the solid red line is the fit to the distribution, while the dashed red line is the extrapolation of the fit. The result of the fit is shown in each plot.}
    \label{fig:exDrho}
\end{figure}

Figure~\ref{fig:exDrho} shows an example of the described method for the case of $m_T \in [0.398, 0.445]$ GeV and $\sqrt{s_{NN}} = 8.8$ GeV.

\subsection{Momentum space method}\label{sec:cq}

An alternative method to the one described in the previous section is to perform the Fourier transform of the pair-distance distribution using a discrete Fourier transform~\cite{FFTW} to obtain its representation in momentum space, where the three-dimensional correlation function takes the analytical form
\begin{equation}
\label{eq:Cq-levy}
    C_2(q_\text{out}, q_\text{side}, q_\text{long}) = 1 + \lambda \, \exp[-(\vec{q}\,^T R^2 \vec{q})^{\flatfrac{\alpha}{2}}],
\end{equation}
with $\vec{q}$ the relative momentum of the pair. Therefore, the same five parameters of the configuration space method are left: $\lambda$, $\alpha$, $R_\text{out}$, $R_\text{side}$, and $R_\text{long}$. The discrete Fourier transform takes the pair-distance distribution with a uniform cubic grid of size N, normalized to unity over the complete grid, that is
\begin{equation}
    \widetilde{D}(\vec{\rho_i}) = \frac{D(\vec{\rho_i})}{\displaystyle{\sum_{j,k,l = 0} ^{N+1} D(\rho_{j,x}, \rho_{k,y}, \rho_{l,z})}}.
\end{equation}
Then, its discrete Fourier transform is
\begin{equation}
    \widetilde{C} (\vec{q}_m) = 1+ \sum_{\vec{n} = 0} ^{N - 1} \widetilde{D}(\vec{\rho}_n)\: e^{\flatfrac{2\pi i \vec{m}\cdot\vec{n}}{N}},
\end{equation}
where $\vec{n} = (n_x, n_y, n_z)$ and $\vec{m} = (m_x, m_y, m_z)$ represent each spatial and momentum grid position, respectively. The momentum grid positions can be related to its physical momentum through

\begin{equation}
    m_i = \begin{cases}
        n & \text{ if } n\leq \flatfrac{(N-1)}{2}, \\
        n - N & \text{ if } n > \flatfrac{(N-1)}{2},
    \end{cases}
\end{equation}
for $i = x,\; y,\; z$. Therefore, the physical momentum component along the $i-$th direction is
\begin{equation}
    q_i = m_i \ \frac{2\pi}{N \Delta\rho_i},
\end{equation}
with $\Delta \rho_i$ the separation between points in the spatial grid. In this way, the momentum values range from negative to positive and are centered around $q_i = 0$. Here, we have used $N = 101$ and $\Delta\rho_i = 1.9802$ fm. The resulting correlation function does not retain the original statistical uncertainties. Therefore, when fitting Eq.~\eqref{eq:Cq-levy}, we rescale the parameter uncertainties by $\sqrt{\flatfrac{\chi^2}{\text{NDF}}}$.

Just as in the configuration space method, the sources of systematic error are the number of analyzed events, $N_\text{events}$, $q_\text{LCMS} ^\text{max}$ and the fitting interval. For the number of events and the maximum relative momentum we take the same values as in the configuration space method, while for the fitting interval we take \mbox{$q_\text{min, max} = \pm\: 0.157, \; \pm \: 0.208, \: \pm\: 0.235$ and $\pm\:0.313$ GeV}, with the default value of \mbox{$q_\text{min, max} ^\text{def} = \pm \:0.313$ GeV.}

As a function of the number of analyzed events, $C_2 (q)$ has fluctuations smaller than 3\% for the considered $N_\text{events}$ values. As a function of $c$, $C_2(q)$ changes by less than 5\% and as a function of $q_\text{min, max}$, the fluctuations are smaller than 1\%.

\begin{figure}
    \centering
    \includegraphics[width=\linewidth]{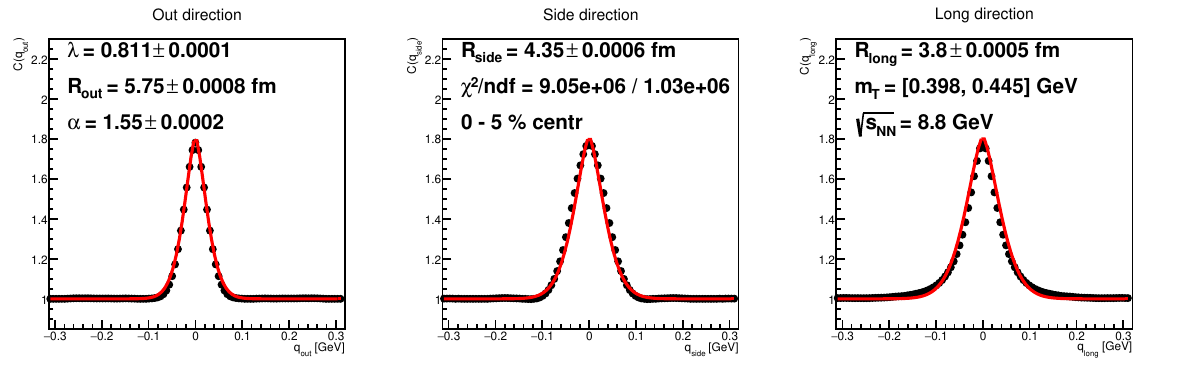}
    \caption{Example of the fitting method in momentum space to the two-pion correlation function for each projection into outward, sideward, and longitudinal directions, obtained from ${}^{209}\mathrm{Bi}+{}^{209}\mathrm{Bi}$ collisions within the EPOS 4.0.3 framework with the 2D-TExS EoS. The black dots are the obtained correlation function, while the red solid line is the obtained fit. The result of the fit is shown in each plot.}
    \label{fig:exCq}
\end{figure}

Figure~\ref{fig:exCq} shows an example of the method in momentum space for $m_T \in [0.398, 0.445]$ GeV and $\sqrt{s_{NN}} = 8.8$ GeV, the same as in Fig.~\ref{fig:exDrho}. As can be seen, there is a discrepancy between the results in Fig.~\ref{fig:exDrho} and those in Fig.~\ref{fig:exCq} . This discrepancy may arise from deviations of the source from a perfectly ellipsoidal (elliptically contoured) Lévy shape, as well as from the different weights assigned to the source points in each method. Since the pair-distance distribution is not a directly measurable quantity, the momentum space method is more directly comparable to experiments.

\section{Results}\label{sec:results}

In this work, we simulated ${}^{209}\mathrm{Bi}+{}^{209}\mathrm{Bi}$ collisions using EPOS 4.0.3 with both hydrodynamical evolution (employing the 2D-TExS EoS), \textit{core-corona} separation and hadronic afterburner turned on. The simulations were made for the following collision energies: $\sqrt{s_{NN}} = 5.8$, 7.1, 7.7, 8.3, 8.8, 9.2, 10.1, and 11 GeV in the centrality class of 0--5\%.

By applying the methods described in Sec.~\ref{sec:dist}, we can explore the collision energy dependence of the five fit parameters and compare the results of both methodologies. 

\begin{figure}
    \centering
    \subfloat[]{\includegraphics[width=0.5\linewidth]{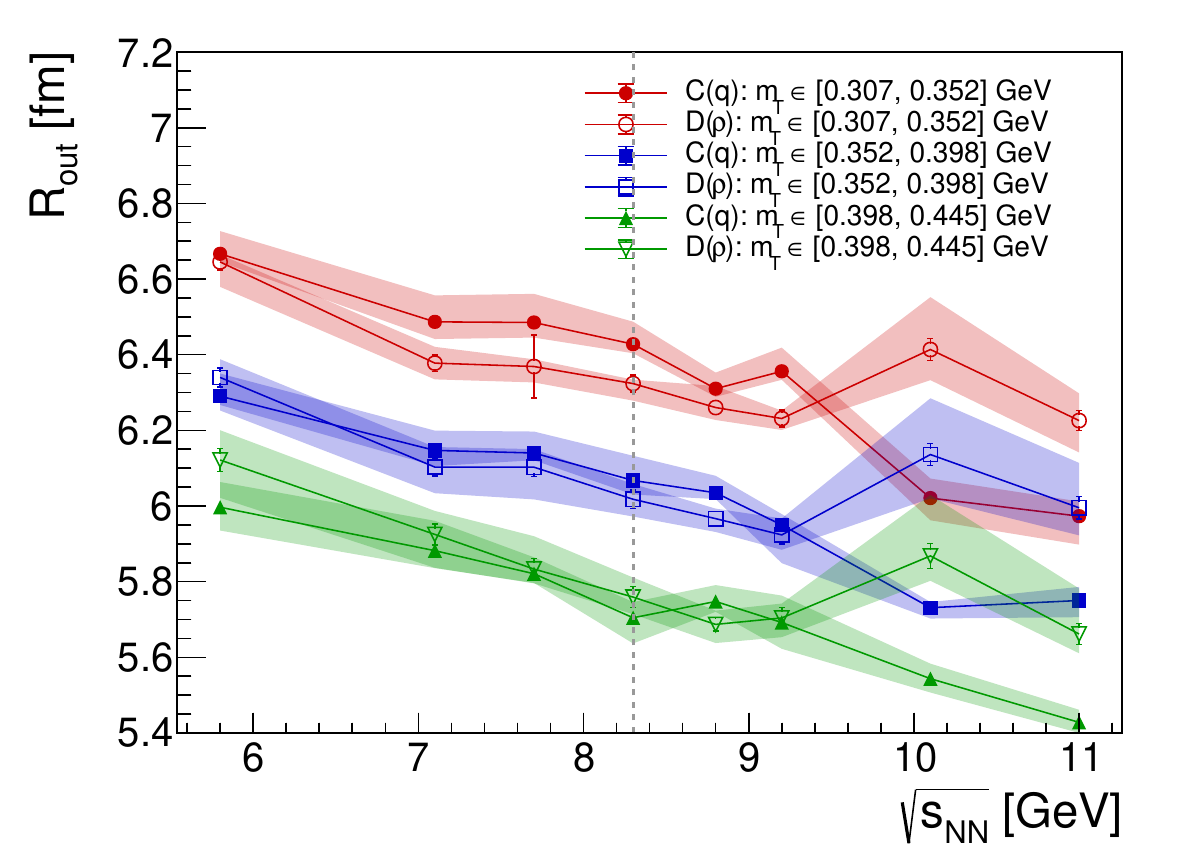}}
    \subfloat[]{\includegraphics[width=0.5\linewidth]{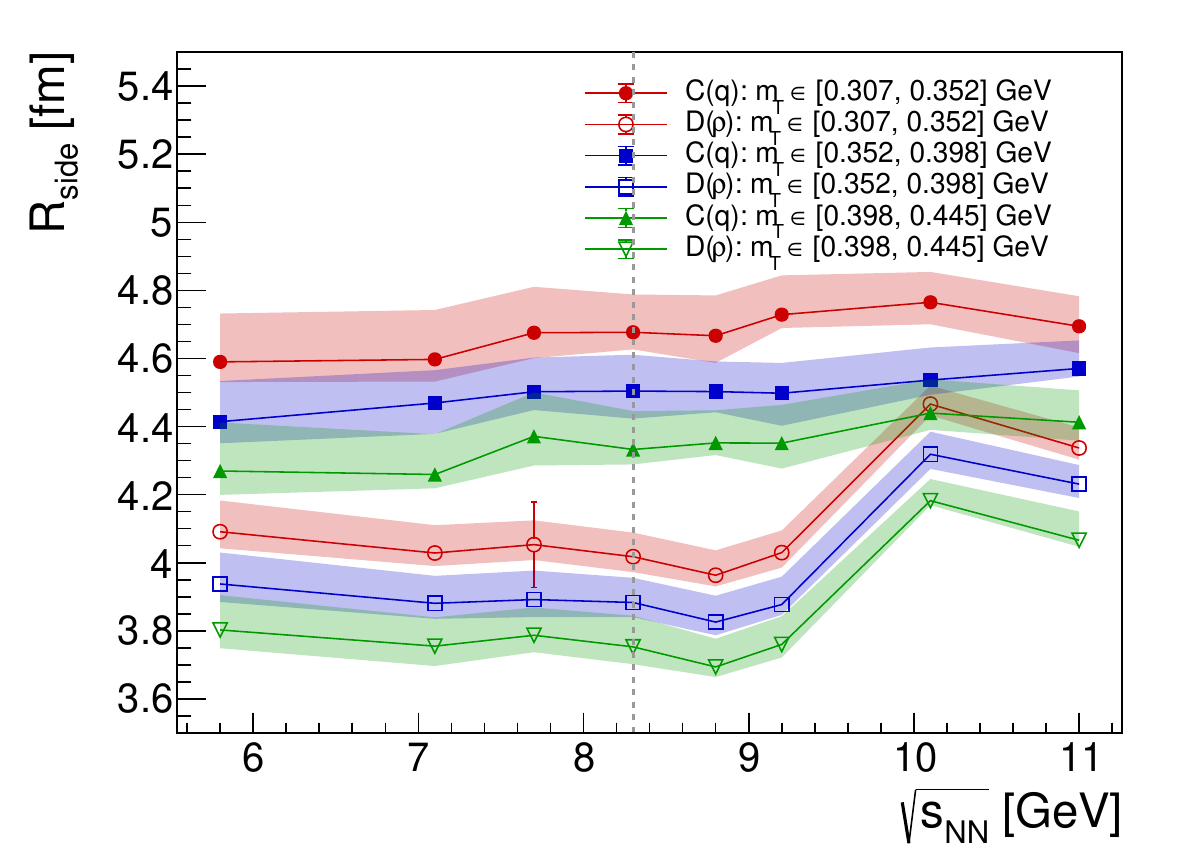}}\\
    \subfloat[]{\includegraphics[width=0.5\linewidth]{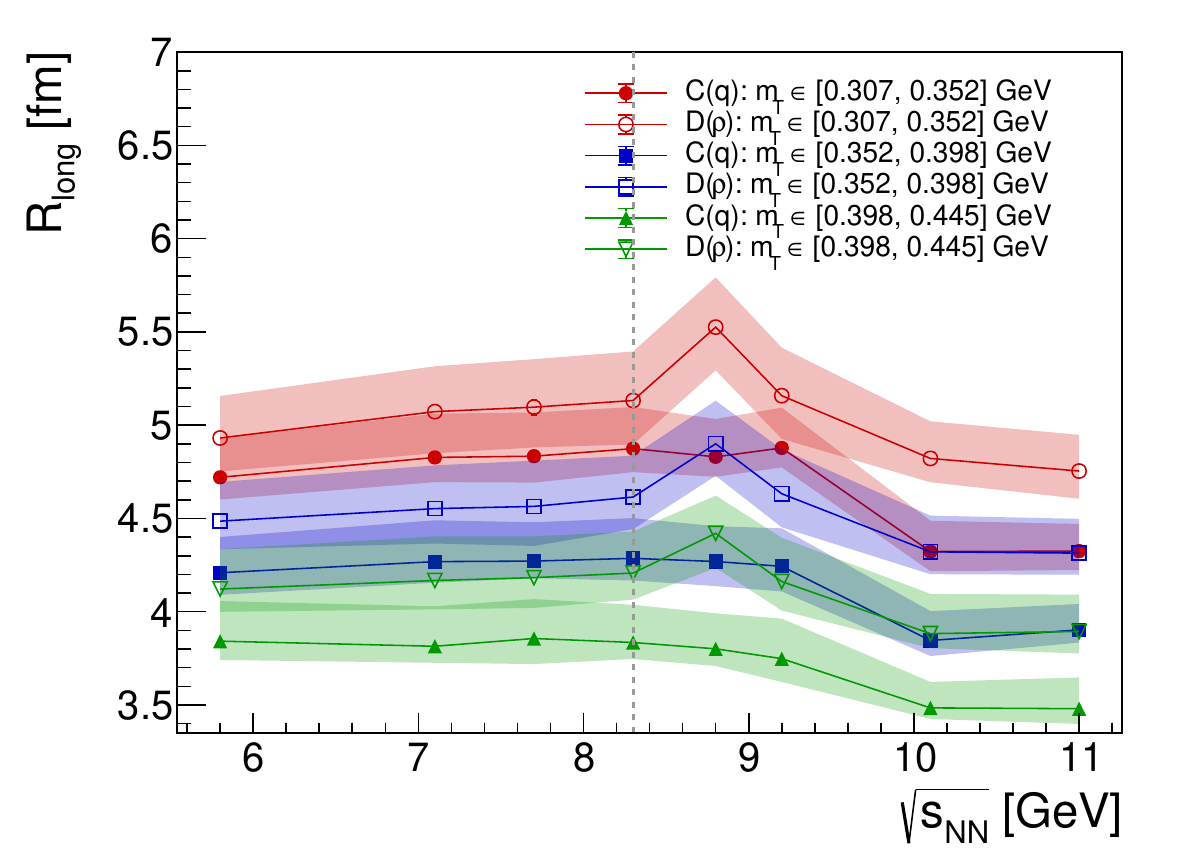}}
    \subfloat[]{\includegraphics[width=0.5\linewidth]{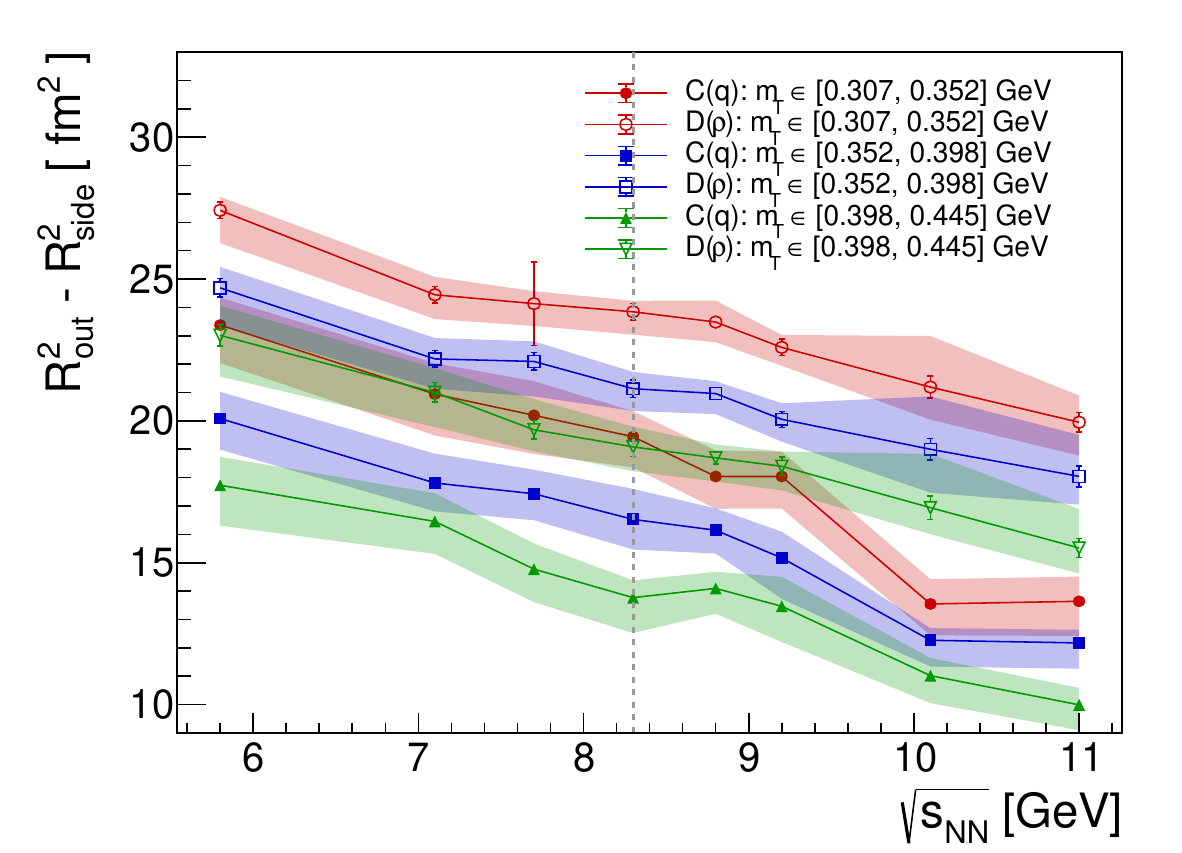}}
    \caption{Dependence of the fit parameters (a) $R_\text{out}$, (b) $R_\text{side}$, (c) $R_\text{long}$ and (d) $R_\text{out}^2 - R_\text{side}^2$ on the collision energy, for different transverse mass intervals. The results shown in empty symbols are obtained from the configuration space method, while the full symbols show the momentum space method. The color bands represent the systematic uncertainty. The vertical dashed gray line represents the energy at which the freeze-out $\mu_B$ equals $\mu_{BC}$.}
    \label{fig:Radii}
\end{figure}

\begin{figure}
    \centering
    \subfloat[]{\includegraphics[width=0.5\linewidth]{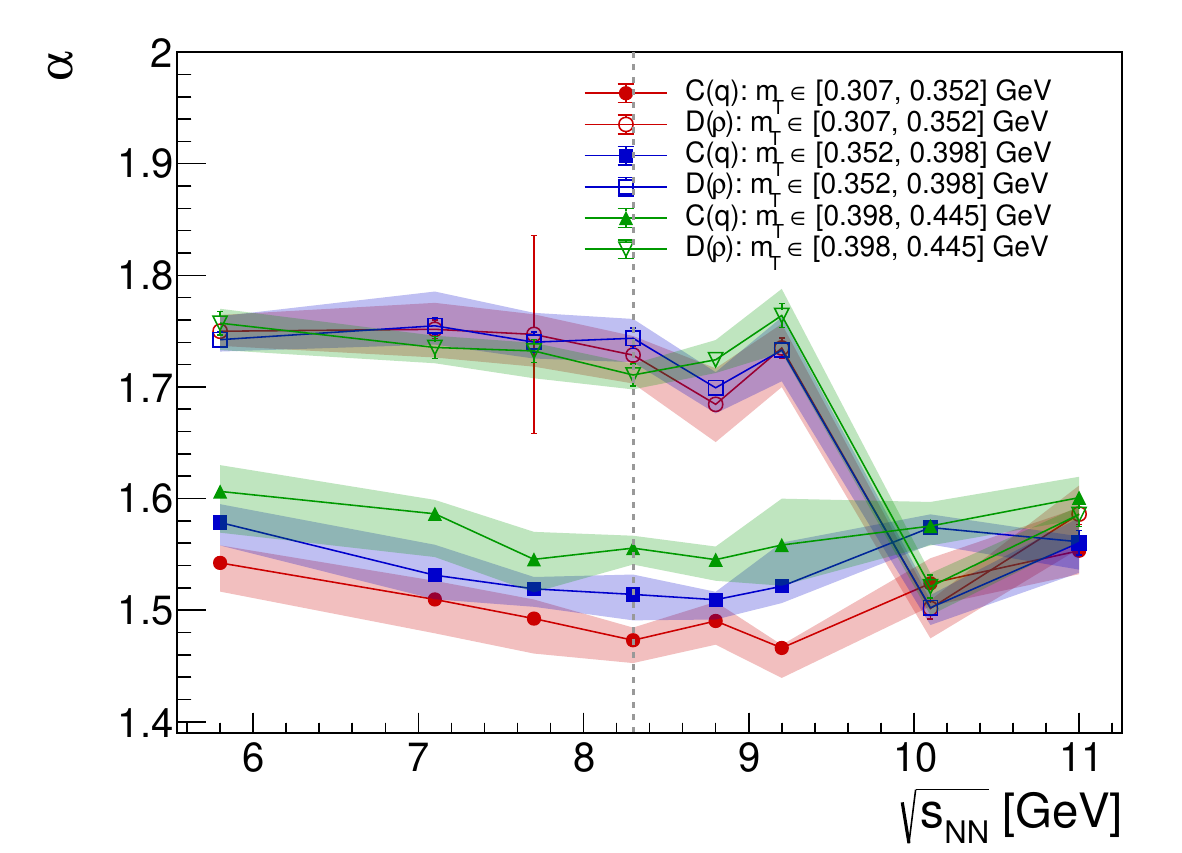}}
    \subfloat[]{\includegraphics[width=0.5\linewidth]{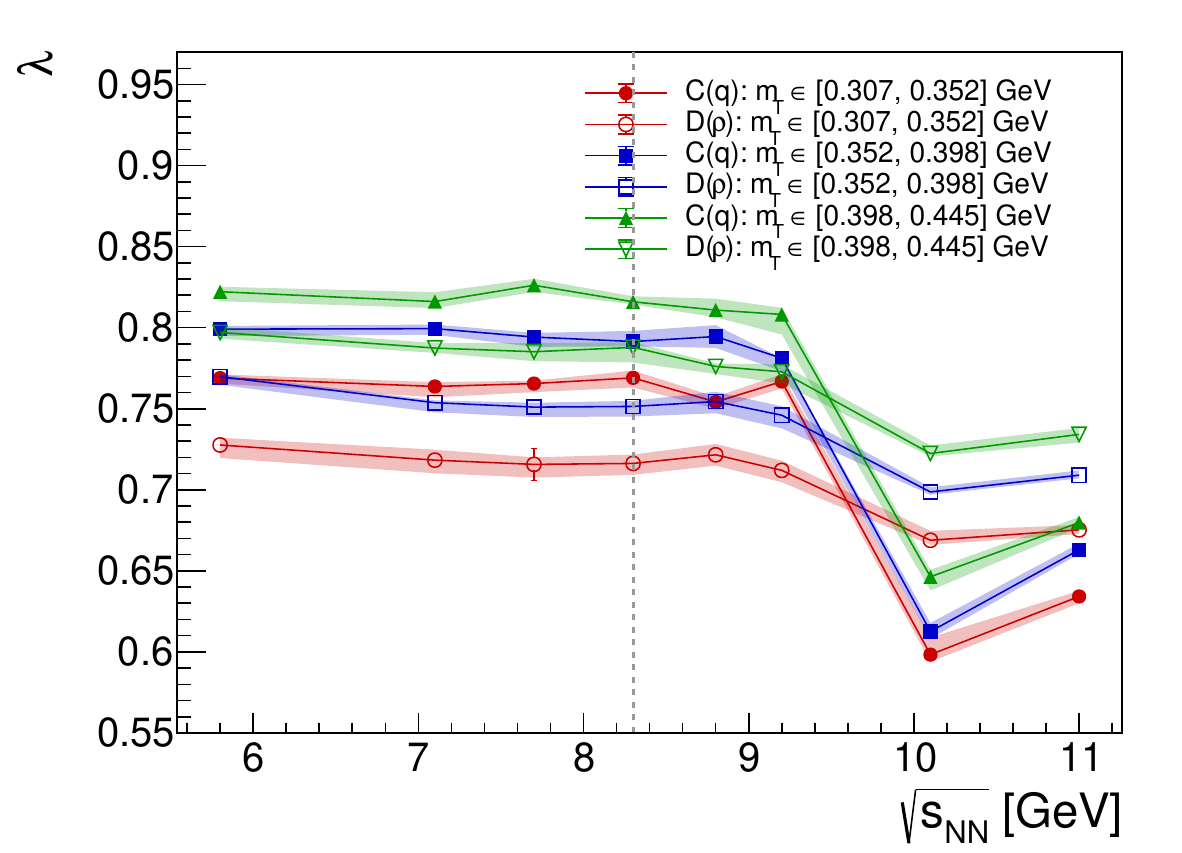}}
    \caption{Dependence of the fit parameters (a) $\alpha$ and (b) $\lambda$ on the collision energy, for different transverse mass intervals. The results shown in empty symbols are obtained from the configuration space method, while the full symbols show the momentum space method. The color bands represent the systematic uncertainty. The vertical dashed gray line represents the energy at which the freeze-out $\mu_B$ equals $\mu_{BC}$.}
    \label{fig:alpha-lambda}
\end{figure}

Figures~\ref{fig:Radii} and~\ref{fig:alpha-lambda} show the dependence of the fit parameters on the collision energy, as well as for the derived quantity $R_\text{out}^2 - R_\text{side}^2$. This last quantity, for the case of Gaussian sources, is related to the emission time; nonetheless, this relation no longer holds for Lévy sources~\cite{Csanad:2024jpy}. However, Refs.~\refcite{Huang:2025edi},~\refcite{Csanad:2024jpy} and~\refcite{Lacey:2014wqa} have highlighted the importance of this quantity as a possible signal of criticality. We also note the difference in the obtained results between both methods. For $R_\text{out}$, $R_\text{side}$ and $\lambda$, the momentum space method yields larger results than the configuration space method for collision energies $\sqrt{s_{NN}} \leq 9.2$ GeV, while for $\sqrt{s_{NN}} \geq 10.1$ GeV, the order reverses. In contrast, for $R_\text{long}$, $\alpha$ and $R_\text{out}^2 - R_\text{side}^2$, the momentum space method yields lower results than the configuration space method, overall. The results are particularly different for the $R_\text{side}$, $\alpha$ and $\lambda$ parameters.

The results in Figs.~\ref{fig:Radii} and~\ref{fig:alpha-lambda} show no dip for $\alpha$ and no clear change for $R_\text{out}^2 - R_\text{side}^2$ that could be interpreted as a signal of criticality. A comparison with a non-critical baseline is necessary to draw firmer conclusions about the possible influence of critical phenomena on these parameters.

\enlargethispage{\baselineskip}
In general, none of the fit parameters show a strong dependence on the collision energy, and all exhibit a change in trend between 9 and 10 GeV. This change is more noticeable for $\lambda$. This behavior may be explained with an increasing contribution from long-lived resonance decays, as suggested by the \textit{core-halo} interpretation of the intercept parameter.
Recall that, in the \textit{core-halo} model~\cite{Bolz:1992hc, Csorgo:1994in} \footnote{Do not confuse the \textit{core-halo} model with the \textit{core-corona} separation, where the \textit{core} of the former is related to the size of region that is resolvable by a detector with finite resolution, while the \textit{core} in the latter is related to a density criterion that modifies the interactions of the particles in the fireball.}, the $\lambda$ parameter is related to the squared fraction of particles coming from the \textit{core} with respect to the total; therefore, such a decrease in $\lambda$ is related to an increase in the number of pions produced by long-lived resonances decaying far away.

Our results can also be compared with the results of Ref.~\refcite{Ayala:2023sbb}, which were obtained with UrQMD~\cite{Bass:1998ca, Bleicher:1999xi} alone, where a decrease in $\lambda$ of around 10\% is observed, while the radius and $\alpha$ remain almost constant. This is in line with the overall findings of the present analysis and further supports the interpretation that the observed change in trend is not related to critical phenomena, but rather to processes associated with the increase in collision energy. Note, however, that the particular values are actually different, due to the different event generators and the use of a hydrodynamical and \textit{core-corona} description of the collision in the EPOS framework. 

\section{Summary and conclusions}\label{sec:conc}

In this work, we have simulated central Bi+Bi collisions within the EPOS 4.0.3 framework and explored the collision energy dependence of the parameters that characterize the two-pion correlation function from the pair-distance distribution, in configuration space and in momentum space. The Monte Carlo simulations included the 2D-TExS equation of state that includes a critical end-point located at a temperature $T_C = 133.36$ MeV and a baryonic chemical potential $\mu_{B,C} = 400$ MeV, which roughly corresponds to a collision energy of approximately $\sqrt{s_{NN}} \approx 8.3$ GeV. We see a difference in the obtained results between the momentum space and the configuration space method, especially for $\sqrt{s_{NN}} \leq 9.2$ GeV. This discrepancy may be attributed to deviations of the source from a perfectly ellipsoidal Lévy shape.

Our results also show a weak dependence of the source parameters on the collision energy up to $\sqrt{s_{NN}} = 9.2$ GeV, while for larger energies, there is a change of trend that we interpret as coming from the decrease in the fraction of pions from the core which suggests that more charged pions are produced by long-lived resonances that decay far away from the central region of the collision. We have not observed a clear signal in any of the parameters that could be attributed to the presence of the critical end-point. Nevertheless, there could still be subtle differences between an equation of state with a critical end-point and another one without it. The comparison with the case of an EoS without a critical end-point as well as with the case where the strength of the critical end-point is modified, is work in progress and will be reported elsewhere. Moreover, in the EPOS4 framework, the EoS only affects the hydrodynamical evolution of the system, but it is not expected for the CEP to have a strong impact on the dynamical evolution of the fireball. Also, the hadronic afterburner does not incorporate the effects of the EoS. In addition, the EPOS4 framework does not account for the expected critical behavior of the transport coefficients, since we expect the transport coefficients to have a non-trivial dependence on the EoS in the critical region. For instance, the bulk and shear viscosities are expected to scale with the correlation length near the critical point. This could have a stronger impact on the pion correlations. A more complete treatment of these effects could potentially reveal clearer signatures of critical phenomena in the observables studied here. This effects will be investigated in future work.

\section*{Acknowledgments}
S.B.-L. acknowledges the financial support of a fellowship granted by
Secretaría de Ciencia, Humanidades, Tecnología e Innovación (SECIHTI) México as part of the Sistema Nacional de Posgrados. S.B.-L. further acknowledges the staff and colleagues at the Institute of Physics and Astronomy, ELTE, for their kind hospitality during a visit, during which part of this work was carried out, and thanks L.A. Hernández and Universidad Autónoma Metropolitana for the financial support for the visit. M. Cs. thanks for the support of NKFIH through grants NKKP ADVANCED 152097 and 2026-4.1.1-MISSZIÓK-2026-00005. D. K. was supported by the NKFIH grant PD-146589. Support for this work has also been received in part by DGAPA-PAPIIT-UNAM grant number AG100826, and from Secretaría de Ciencia, Humanidades, Tecnología e Innovación (SECIHTI) México grant numbers CIORGANISMOS-2025-17, CBF-2025-G-1718, and CBF-2026-465.

\bibliographystyle{ws-ijmpa}
\bibliography{sample}
\end{document}